\documentclass[preprint,preprintnumbers, prd, floatfix,  superscriptaddress,nofootinbib] {revtex4-1}
\usepackage{graphicx}
\usepackage{epsfig}
\usepackage{bm}
\usepackage{amssymb}
\usepackage{float}
\usepackage{amsmath}
\usepackage{dcolumn}
\usepackage{cancel}
\usepackage[colorlinks]{hyperref}
\usepackage[usenames,dvipsnames]{color}
\usepackage{color}
\usepackage{epstopdf}
\usepackage{nccbbb}
\hypersetup{
     breaklinks=true,
    pdfstartview={FitH},    colorlinks=true,       
    linkcolor=blue,          
    citecolor=red,        
    filecolor=magenta,      
    urlcolor=blue,           
    anchorcolor=green,      
    linktocpage=true
}
\begin{document}

\title {A new coupled three-form dark energy model and implications for the $H_0$ tension}

\author{Yan-Hong Yao}
\email{yhy@mail.nankai.edu.cn}
\author{Xin-He Meng}
\email{xhm@nankai.edu.cn}

\affiliation{Department of Physics, Nankai University, Tianjin 300071, China}

\begin{abstract}
We propose a new coupled three-form dark energy model to relieve the Hubble tension in this paper. Firstly, by performing a dynamical analysis with the coupled three-form dark energy model, we obtain four fixed points, including a saddle point representing a radiation dominated Universe, a saddle point representing a matter dominated Universe, and two attractors representing two saturated de Sitter Universes. Secondly, by confronting the coupled three-form dark energy model and the $\Lambda$ cold dark matter model (the $\Lambda$CDM model) with cosmic microwave background (CMB), baryonic acoustic oscillations (BAO), Type Ia supernovae (SN Ia) observations, we obtain $H_0= 67.8_{-0.6}^{+0.7}$($1\sigma$ level) km/s/Mpc for the coupled three-form dark energy model and $H_0=67.6_{-0.5}^{+0.5}$($1\sigma$ level) km/s/Mpc for the $\Lambda$CDM model, the former is in strong tension with the latest local measured $H_0$ value at $4.3\sigma$ confidence level, while the latter is in strong tension with the latest local measured $H_0$ value at $5.1\sigma$ level.

\textbf{}
\end{abstract}

\maketitle

\section{Introduction}
\label{intro}
The dominant components of the Universe, dark energy and dark matter, are incorporated in the standard model of cosmology, known as the $\Lambda$CDM model. In such model, dark energy takes the form of a cosmological constant $\Lambda$. Meanwhile, the dark matter is nonrelativistic, and it interacts with ordinary matter only through gravity.

Although mathematically simple, the $\Lambda$CDM model provides an excellent fit to a wide range of cosmological data. However, an exception is emerging in the Hubble constant $H_0$. In 2018, the Planck satellite measured a $H_0$ value of $67.4\pm0.5$($1\sigma$ level) km/s/Mpc from a $\Lambda$CDM fit to the CMB \cite{aghanim2018planck}. In 2019, The SH0ES collaboration yielded a latest $H_0$ value of
$74\pm1.4$($1\sigma$ level) km/s/Mpc from direct measurements by using so-called standard candles: type Ia supernovae and Cepheid variable stars \cite{riess2019large}. Recently, the H0LiCOW collaboration obtained a $H_0$ value of $73.3_{-1.8}^{+1.7}$($1\sigma$ level) km/s/Mpc by using gravitationally lensed quasar \cite{wong2019h0licow}. Combining the SH0ES
and H0LiCOW measurements gives a model independent $H_0$ value of $73.8\pm1.1$($1\sigma$ level) km/s/Mpc, which is in 5.3$\sigma$ tension with the  $\Lambda$CDM prediction.

Since preliminary attempt to resolve the $H_0$ tension by searching considerable systematic errors in the Planck observation and the local measurements have failed \cite{spergel2015planck,addison2016quantifying,aghanim2017planck,cardona2017determining,follin2018insensitivity}, increasing attention is focusing on the possibility that the $\Lambda$CDM model is not the final picture. For example, ref.\cite{Li2013Planck,huang2016how} pointed out that a phantom dark energy prefers a high value of $H_0$. Refs. \cite{battye2014evidence,Zhang2014Neutrinos,zhang2015sterile,feng2018searching,zhao2018measuring,choudhury2019constraining} showed that considering extra relativistic degrees of freedom $N_{eff}$ in the $\Lambda$CDM model favors a high value of $H_0$ when $N_{eff}>3.046$. Although phantom dark energy and extra relativistic species can help with the $H_0$ tension, it is worth to mention that these solutions are disfavored from both BAO and SN Ia data and from a model comparison point of view. \cite{vagnozzi2020new} Refs. \cite{agrawal2019rock,poulin2019early} introduced an exotic early dark energy (EDE) that acts as a cosmological constant before a critical redshift $z_c$ around 3000 but whose density then dilutes faster than radiation to resolve the $H_0$ tension. Ref.\cite{li2019simple} introduced a emergent dark energy with its equation of state increases from $-\tfrac{2}{3\mathrm{ln}\,10}-1$ in the past to $-1$ in the future to handle the Hubble problem. Ref.\cite{2018Vacuum} showed that the Parker vacuum metamorphosis (VM) model, physically motivated by quantum gravitational effects, can remove the Hubble tension. In addition, an interacting dark energy can affect the constraint results of $H_0$, which provides another way to address the Hubble constant problem.\cite{di2017can,yang2018interacting}

In Ref.\cite{yao2018a}, we proposed a power-law coupled three-form dark energy model and successfully alleviated the coincidence problem with this model. Agreeing with the opinion in \cite{yao2018a} that dark energy might be represented by a three-form field, and considering the fact that interacting dark energy affects the constraint results of $H_0$, in this paper, we put forward a
new coupled three-form dark energy model to relieve the Hubble tension.

The contents of this paper are as follows. In section \ref{sec:1}, we consider a coupled dark energy model in which dark energy is represented by a three-form field and other components are represented by ideal fluids. In section \ref{sec:2}, we derive the autonomous system of evolution equations, and analyze the stability of its fixed points. In section \ref{sec:3}, we confront the model with the data from CMB, BAO, SN Ia observations. In the last section, we make a brief conclusion with this paper. For convenience, we set 100 km/s/Mpc=1,i.e., $H_0=h$ in the following part of the paper.

\section{A coupled three-form dark energy model}
\label{sec:1}
In this section, a new coupled three-form dark energy model is presented, in which dark energy is represented by a three-form field and other components are represented by ideal fluids. We restrict the coupling here to be the conformal form \cite{koivisto2013coupled,yao2018a}, a case that has been thoroughly studied in the context of scalar fields. The total Lagrangian is written as
\begin{equation}
\mathcal{L}=\frac{R}{2\kappa^2}+\mathcal{L}_{(d)}+\mathcal{L}_{(b)}+\mathcal{L}_{(\gamma)}+\mathcal{L}_{(\nu)},
\end{equation}
where $R$ denotes the Ricci scalar and $\kappa=\sqrt{8\pi G}$ is the inverse of the reduced Planck mass.
$d$, $b$, $\gamma$, and $\nu$ denote dark sectors, baryon, photon, and neutrino, respectively. For ideal fluids, the derivation of the  equations of motion from a variational principle is complicated since one needs to consider the constraint equations satisfied by the fluid variables. To solve this problem, several variational formulations have been proposed. In this paper, we consider the variational formulations discussed in \cite{ray1972lagrangian}, then each Lagrangian can be expressed as
\begin{eqnarray}
\mathcal{L}_{(d)} &=& -\frac{1}{48}F^{2}-I(A^{2})\tilde{\rho}_{(c)}+\lambda_{1}(g_{\mu\nu}u_{(c)}^{\mu}u_{(c)}^{\nu}+1)+\lambda_{2}\nabla_{\alpha}(\tilde{\rho}_{(c)}u_{(c)}^{\alpha}), \\
\mathcal{L}_{(b)} &=&-\rho_{(b)}+\lambda_{3}(g_{\mu\nu}u_{(b)}^{\mu}u_{(b)}^{\nu}+1)+\lambda_{4}\nabla_{\alpha}(\rho_{(b)}u_{(b)}^{\alpha}), \\
\mathcal{L}_{(\gamma)} &=&
-\tilde{\rho}_{(\gamma)}(1+\epsilon_{(\gamma)}(\tilde{\rho}_{(\gamma)}))+\lambda_{5}(g_{\mu\nu}u_{(\gamma)}^{\mu}u_{(\gamma)}^{\nu}+1)+\lambda_{6}\nabla_{\alpha}(\tilde{\rho}_{(\gamma)}u_{(\gamma)}^{\alpha}), \\
\mathcal{L}_{(\nu)} &=& -\tilde{\rho}_{(\nu)}(1+\epsilon_{(\nu)}(\tilde{\rho}_{(\nu)}))+\lambda_{7}(g_{\mu\nu}u_{(\nu)}^{\mu}u_{(\nu)}^{\nu}+1)+\lambda_{8}\nabla_{\alpha}(\tilde{\rho}_{(\nu)}u_{(\nu)}^{\alpha}),
\end{eqnarray}
$A$ and $F=dA$ represent the three-form field and the field strength tensor, $I(A^2)$ is referred to as coupling function. $\tilde{\rho}$ denotes the rest density, and $\epsilon$ denotes the rest, specific internal energy, which is connected with pressure through the following relation,
\begin{eqnarray}
  \frac{d\epsilon}{d\tilde{\rho}}&=& \frac{p}{\tilde{\rho}^{2}}
\end{eqnarray}
Finally, $\lambda_1-\lambda_8$ are the multipliers.

One now can obtain the field equations from the total action
\begin{equation}\label{}
 S=\int\mathcal{L}\sqrt{-g}d^{4}x.
\end{equation}
The variation of the action with respect to the $g^{\mu\nu},\tilde{\rho}_{(c)},u_{(c)}^{\alpha},\rho_{(b)},u_{(b)}^{\alpha},\tilde{\rho}_{(\gamma)},u_{(\gamma)}^{\alpha},\tilde{\rho}_{(\nu)},u_{(\nu)}^{\alpha}$ leads to
\begin{eqnarray}
R_{\mu\nu}-\frac{1}{2}g_{\mu\nu}R &=& \kappa^{2}T_{\mu\nu}, \\
(\partial_{\alpha}\lambda_{2})u_{(c)}^{\alpha} &=& -I, \\
\lambda_{1} &=& \frac{1}{2}I\tilde{\rho}_{(c)}, \\
(\partial_{\alpha}\lambda_{4})u_{(b)}^{\alpha} &=& -1, \\
\lambda_{3} &=& \frac{1}{2}\rho_{(b)}, \\
(\partial_{\alpha}\lambda_{6})u_{(\gamma)}^{\alpha} &=& -(1+\epsilon_{(\gamma)}(\tilde{\rho}_{(\gamma)})),\\
\lambda_{5} &=& \frac{1}{2}(1+\epsilon_{(\gamma)}(\tilde{\rho}_{(\gamma)}))\tilde{\rho}_{(\gamma)}, \\
(\partial_{\alpha}\lambda_{8})u_{(\nu)}^{\alpha} &=& -(1+\epsilon_{(\nu)}(\tilde{\rho}_{(\nu)})), \\
\lambda_{7} &=& \frac{1}{2}(1+\epsilon_{(\nu)}(\tilde{\rho}_{(\nu)}))\tilde{\rho}_{(\nu)},
\end{eqnarray}
with the help of equation (6), and $(9)-(16)$, the total energy-momentum tensor for three-form field and ideal fluids is written as
\begin{equation}\label{}
\begin{split}
 T_{\mu\nu}=&\frac{1}{6}F_{\mu\alpha\beta\gamma}F_{\nu}^{\alpha\beta\gamma}+6\frac{dlnI}{dA^{2}}\rho_{(c)}A_{\mu}^{\alpha\beta}A_{\nu\alpha\beta}-g_{\mu\nu}\frac{1}{48}F^{2}\\
 &+\rho_{(c)}u_{(c)\mu}u_{(c)\nu}+\rho_{(b)}u_{(b)\mu}u_{(b)\nu}+(\rho_{(\gamma)}+p_{(\gamma)})u_{(\gamma)\mu}u_{(\gamma)\nu}+p_{(\gamma)}g_{\mu\nu}+
 (\rho_{(\nu)}+p_{(\nu)})u_{(\nu)\mu}u_{(\nu)\nu}+p_{(\nu)}g_{\mu\nu},
\end{split}
\end{equation}
where $\rho_{(c)}=I(A^{2})\tilde{\rho}_{(c)}$, $\rho_{(\gamma)}=\tilde{\rho}_{(\gamma)}(1+\epsilon_{(\gamma)}(\tilde{\rho}_{(\gamma)}))$, and $\rho_{(\nu)}=\tilde{\rho}_{(\nu)}(1+\epsilon_{(\nu)}(\tilde{\rho}_{(\nu)}))$.
Since we assume that the Universe is homogeneous and isotropic, we have $u_{\mu}=u_{(c)\mu}=u_{(b)\mu}=u_{(\gamma)\mu}=u_{(\nu)\mu}$. In addition, we denote $\rho_{(r)}=\rho_{(\gamma)}+\rho_{(\nu)}$ and $p_{(r)}=p_{(\gamma)}+p_{(\nu)}$. As a result, (17) becomes to
\begin{equation}\label{}
\begin{split}
 T_{\mu\nu}=&\frac{1}{6}F_{\mu\alpha\beta\gamma}F_{\nu}^{\alpha\beta\gamma}+6\frac{dlnI}{dA^{2}}\rho_{(c)}A_{\mu}^{\alpha\beta}A_{\nu\alpha\beta}-g_{\mu\nu}\frac{1}{48}F^{2}\\
 &+\rho_{(c)}u_{\mu}u_{\nu}+\rho_{(b)}u_{\mu}u_{\nu}+(\rho_{(r)}+p_{(r)})u_{\mu}u_{\nu}+p_{(r)}g_{\mu\nu}.
\end{split}
\end{equation}
By varying the total action with respect to the three-form field, we have the following equation of motion,
\begin{equation}\label{}
  \nabla_{\alpha}F^{\alpha\mu\nu\rho} = 12\frac{dlnI}{dA^{2}}\rho_{(c)}A^{\mu\nu\rho}.
\end{equation}
Using the equation of motion for the three-form field and the condition that the divergence
of the total stress energy tensor vanishes, we have the equation of motion for the dark matter:
\begin{equation}\label{}
  \nabla_{\mu}(\rho_{(c)}u^{\mu}u_{\nu})=-2\frac{dlnI}{dA^{2}}\rho_{(c)}A^{\alpha\beta\gamma}\nabla_{\nu}A_{\alpha\beta\gamma}.
\end{equation}
The homogeneous, isotropic, and spatially flat space-time is described by the Friedmann-
Robertson-Walker (FRW) metric,
\begin{equation}\label{}
  ds^{2}=-dt^{2}+a(t)^{2}d\vec{x}^{2},
\end{equation}
and the three-form field is assumed as a time-like component of a dual vector
field in order to be compatible with FRW symmetries \cite{Koivisto2009Inflation2}.
\begin{equation}
  A_{i j k}=X(t)a(t)^{3}\varepsilon_{ijk}.
\end{equation}
To specify a coupled three-form dark energy model, we set the coupling function $I$ to be
\begin{equation}\label{}
  I=(1+\frac{\kappa^{2}}{6}A^{2})^{\frac{\lambda}{2}}=(1+\kappa^{2}X^{2})^{\frac{\lambda}{2}},
\end{equation}
where $\lambda$ is the coupling constant. Function $I$ is selected without any fundamental reason.
Now we have the Friedmann equations:
\begin{eqnarray}
  H^{2} &=& \frac{\kappa^{2}}{3}\rho, \\
  \dot{H} &=&-\frac{\kappa^{2}}{2}(\rho+p),
\end{eqnarray}
where
\begin{eqnarray}
  \rho &=&- T_{0}^{0}=-g^{00}T_{00}=\frac{1}{2}(\dot{X}+3HX)^{2}+\rho_{(c)}+\rho_{(b)}+\rho_{(r)}, \\
  p &=&\frac{1}{3}T_{i}^{i}=\frac{1}{3}g^{ii}T_{ii}=-\frac{1}{2}(\dot{X}+3HX)^{2}+\lambda\frac{\kappa^{2}X^{2}}{1+\kappa^{2}X^{2}} \rho_{(c)}+\frac{1}{3}\rho_{(r)}.
\end{eqnarray}
The independent equation of motion of the three-form field is
\begin{equation}\label{}
  \ddot{X}+3\dot{H}X+3H\dot{X}+\lambda\frac{\kappa^{2}X}{1+\kappa^{2}X^{2}} \rho_{(c)}=0,
\end{equation}
and the energy conservation equations of two dark sectors is written as
\begin{eqnarray}
  \dot{\rho}_{(c)}+3H\rho_{(c)} &=& \delta H\rho_{(c)},\\
   \dot{\rho}_{(X)}+3H(\rho_{(X)}+p_{(X)})& = &-\delta H\rho_{(c)},
\end{eqnarray}
where
\begin{eqnarray}
 \delta&=&\lambda\frac{\kappa^{2}XX^{\prime}}{1+\kappa^{2}X^{2}}, \\
\rho_{(X)} &=&\frac{1}{2}(\dot{X}+3HX)^{2},\\
p_{(X)} &=& -\frac{1}{2}(\dot{X}+3HX)^{2}+\lambda\frac{\kappa^{2}X^{2}}{1+\kappa^{2}X^{2}} \rho_{(c)}.
\end{eqnarray}

\section{Dynamical system of the coupled three-form dark energy model}
\label{sec:2}
In order to study the dynamical behaviors of the coupled three-form dark energy model, it is convenient to introduce the
following dimensionless variable \cite{Koivisto2009Inflation2}
\begin{equation}\label{}
  x_{1}=\kappa X, \hspace{1cm} x_{2}=\frac{\kappa}{\sqrt{6}}(X^{\prime}+3X), \hspace{1cm} x_{3}=\frac{\kappa \sqrt{\tilde{\rho}_{(c)}}}{\sqrt{3}H}, \hspace{1cm} x_{4}=\frac{\kappa \sqrt{\rho_{(b)}}}{\sqrt{3}H}.
\end{equation}
The autonomous system of evolution equations then can be written as follows by applying the Friedmann equations and equation of motion,
\begin{eqnarray}
x_{1}^{\prime}&=&\sqrt{6}x_{2}-3x_{1},\\
x_{2}^{\prime}&=& (\frac{3}{2}(1+\frac{\lambda x_{1}^{2}}{1+x_{1}^{2}})(1+x_{1}^{2})^{\frac{\lambda}{2}}x_{3}^{2}+\frac{3}{2}x_{4}^2+2(1-x_{2}^2-(1+x_{1}^{2})^{\frac{\lambda}{2}}x_{3}^2-x_{4}^2))x_{2}
 -\frac{\sqrt{6}}{2} \frac{\lambda x_{1}}{1+x_{1}^{2}}(1+x_{1}^{2})^{\frac{\lambda}{2}}x_{3}^{2}, \hspace{1cm} \\
x_{3}^{\prime}&=& -\frac{3}{2}x_{3}+(\frac{3}{2}(1+\frac{\lambda x_{1}^{2}}{1+x_{1}^{2}})(1+x_{1}^{2})^{\frac{\lambda}{2}}x_{3}^{2}+\frac{3}{2}x_{4}^2+2(1-x_{2}^2-(1+x_{1}^{2})^{\frac{\lambda}{2}}x_{3}^2-x_{4}^2))x_{3},\\
x_{4}^{\prime}&=& -\frac{3}{2}x_{4}+(\frac{3}{2}(1+\frac{\lambda x_{1}^{2}}{1+x_{1}^{2}})(1+x_{1}^{2})^{\frac{\lambda}{2}}x_{3}^{2}+\frac{3}{2}x_{4}^2+2(1-x_{2}^2-(1+x_{1}^{2})^{\frac{\lambda}{2}}x_{3}^2-x_{4}^2))x_{4}.
\end{eqnarray}
Here and in the following, the prime stands for the derivative with respect to e-folding time.
\begin{table}[t]
\begin{center}
  \begin{tabular}{|c|c|c|c|c|c|c|c|c|}
\hline
&  & &&&&&&\\[-.5em]
&$(x_1,x_2,x_3,x_4)$& Existence&$\Omega_{(X)}$&$\Omega_{(c)}$&$\Omega_{(b)}$&$\Omega_{(r)}$&$w_{(X)}$&$w_{eff}$
\\[.5em]
\hline
& & & & &&&& \\[-.5em]
(a)&$(0,0,0,0)$&All&$0$&$0$&0&$1$&$-$&$\frac{1}{3}$
\\[.5em]
\hline
& & & & & &&& \\[-.5em]
(b)&$(0,0,x_3,\sqrt{1-x_3^{2}})$&$ 0\leq x_3\leq 1$&$0$&$x_3^{2}$&$1-x_3^{2}$&0&$-$&$0$
\\[.5em]
\hline
& & & & &&&& \\[-.5em]
(c)&$(\sqrt{\frac{2}{3}},1,0,0)$&All&$1$&$0$&$0$&$0$&$-1$&$-1$
\\[.5em]
\hline
& & & & &&&& \\[-.5em]
(d)&$(-\sqrt{\frac{2}{3}},-1,0,0)$&All&$1$&$0$&$0$&$0$&$-1$&$-1$
\\[.5em]
\hline
\end{tabular}
    \caption{Fixed points of the autonomous system.}
\label{fixedpoints}
\end{center}
\end{table}

There are four fixed points for the autonomous system of evolution equations, which are presented in Tab.\ref{fixedpoints}. Fixed point (a) represents a radiation dominated Universe, it is a saddle point since its eigenvalues are
\begin{equation}
  \mu_{(a)}=-3,2,\frac{1}{2},\frac{1}{2}.
\end{equation}
Fixed point (b) represents a matter dominated Universe, its eigenvalues are
\begin{equation}
  \mu_{(b)}=0,-1,\frac{-3-\sqrt{3}\sqrt{27-16 x_{3}^{2}\lambda}}{4},\frac{-3+\sqrt{3}\sqrt{27-16 x_{3}^{2}\lambda}}{4},
\end{equation}
the eigenvector corresponding to the vanishing eigenvalue reads $(0,0,1,\frac{-x_3}{\sqrt{1-x_3^{2}}})$. Generally speaking, we need to go
to the higher order to study the stability of this fixed point, however, as long as one can prove that one of the eigenvalues is positive, in other words, if $\lambda<\frac{3}{2 x_3^2}$ is proved to be true, such fixed point is a saddle point. Since $x_3\in[0,1]$, we only need to prove that $\lambda<\frac{3}{2}$, fortunately, this condition is satisfied, in fact, we will show that $\lambda\ll1$ in the next section.

Fixed point (c) and fixed point (d) are represented by two saturated de Sitter Universes, they are always stable since their eigenvalues both read
\begin{eqnarray}
  \mu_{(c)} &=& -4,-3,-\frac{3}{2},-\frac{3}{2}, \\
  \mu_{(d)} &=&-4,-3,-\frac{3}{2},-\frac{3}{2}.
\end{eqnarray}
One should note that the phase space is separated in two parts because of these two symmetrical attractors, more specifically, generally speaking, the trajectories in the phase space run toward the de Sitter attractor (c) if $x_2>0$ at the beginning, otherwise they will run toward the other attractor.

\section{Confront the model with observations}
\label{sec:3}
In this section, we confront the coupled three-form dark energy model with CMB, BAO, SN Ia observation, and OHD based on the Hubble parameter,
\begin{equation}
 H^2 =\frac{\omega_{(r)}(1+z)^{4}+\omega_{(c)}\left(\frac{1+x_{1}^{2}}{1+x_{10}^{2}}\right)^{\frac{\lambda}{2}}(1+z)^{3}+\omega_{(b)}(1+z)^{3}}{1-x_2^2}.
\end{equation}

\subsection{CMB measurements}
There are two shift parameters, $R$ and $l_{A}$, that contain much of information of CMB power spectrum, the former is defined as
\begin{equation}
  R=\sqrt{\omega_{(m)}}(1+z_{\ast})d_{A}(z_{\ast}),
\end{equation}
and the latter reads
\begin{equation}
  l_{A}=\pi(1+z_{\ast})\frac{d_{A}(z_{\ast})}{r_{s}(z_{\ast})},
\end{equation}
where $d_{A}(z_{\ast})=\frac{1}{1+z}\int_{0}^{z_{\ast}}\frac{d\tilde{z}}{H}$ is the angular distance at decoupling \cite{Efstathiou2010Cosmic}, which depends on the dominant components after decoupling.
The redshift at decoupling $z_{\ast}$ is given by \cite{Hu1996Small}
\begin{eqnarray}
  z_{*}&=&1048(1+0.00124\omega_{(b)}^{-0.738})(1+g_{1}\omega_{(m)}^{g_{2}}),\\
g_{1}&=&\frac{0.0783\omega_{(b)}^{-0.238}}{1+39.5\omega_{(b)}^{0.763}},\\
  g_{2}&=&\frac{0.56}{1+21.1\omega_{(b)}^{1.81}},
\end{eqnarray}
where $\omega_{(m)}=\omega_{(c)}+\omega_{(b)}$.
In this work, we use the following Planck 2018 compressed likelihood \cite{chen2019distance} with these two shift parameters to perform a likelihood analysis,
\begin{eqnarray}
  \chi_{CMB}^{2}&=&s^{T} C_{CMB}^{-1} s,\\
  s &=& (R-1.7502,l_{A}-301.471,\omega_{(b)}-0.02236),
   \end{eqnarray}
where $C_{ij}=D_{ij}\sigma_i\sigma_j$ is the covariance matrix, $\sigma=(0.0046,0.09,0.00015)$ is the errors, and
$D_{CMB}=  \left(
      \begin{array}{ccc}
        1 & 0.46 &-0.66 \\
       0.46 & 1 & -0.33 \\
       -0.66&-0.33& 1\\
      \end{array}
    \right)$ is the covariance.
\subsection{Baryon acoustic oscillations}
The relative BAO distance is defined as
\begin{equation}
   d_{z}(z)=\frac{r_{s}(z_{d})}{D_{V}(z)},
\end{equation}
with $D_{V}(z)=[\frac{z(1+z)^{2}d_{A}(z)^{2} }{H(z)}]^{\frac{1}{3}}$. $z_d$ is the redshift at the drag epoch, a epoch when baryons are released from the Compton drag of the photons. $z_d$ can be calculated by using \cite{Eisenstein1997Baryonic}
\begin{eqnarray}
     z_{d}&=&1291\frac{\omega_{(m)}^{0.251}}{1+0.659\omega_{(m)}^{0.828}}(1+b_{1}\omega_{(b)}^{b_{2}}),\\
     b_{1}&=&0.313\omega_{(m)}^{-0.419}(1+0.607\omega_{(m)}^{0.674}),\\
     b_{2}&=&0.238\omega_{(m)}^{0.223}.
   \end{eqnarray}
We use four measurements from 6dFGS at $z_{eff}$= 0.106, the recent SDSS main
galaxy (MGS) at $z_{eff}$ = 0.15 \cite{ross2015clustering}, and $z_{eff}$= 0.32 and 0.57 for the Baryon Oscillation Spectroscopic Survey (BOSS). \cite{Anderson2013The}.\footnote{Although the BOSS DR12 sample \cite{alam2017clustering} is available now, fitting results won't be too much different if we only use the old data.}
Therefore, the BAO likelihood is written as
\begin{eqnarray}
\chi_{BAO}^{2}&=&t^{T} C_{BAO}^{-1} t,\\
t &=&(d_{z}(0.106)-0.336,d_{z}(0.15)-0.2239,d_{z}(0.32)-0.1181,d_{z}(0.57)-0.0726),
\end{eqnarray}
with $C_{BAO}^{-1}=diag(4444.44,14067.88,183275.76,2004606.79)$ is the the covariance matrix.
\subsection{Type Ia supernovae}
For supernovae data, we employ the Joint Light-curve Analysis (JLA) sample \cite{betoule2014improved},\footnote{For the same reason with giving up new BAO data, we adopt JLA sample in this paper, although the Pantheon sample \cite{scolnic2018complete} is available now.} the distance
modulus is then assumed as
\begin{equation}
  \mu_{obs}=m_B-(M_B-\alpha X_1+\beta c),
\end{equation}
where $m_B$ and $M_B$ are SN Ia peak apparent magnitude and SN Ia absolute magnitude, respectively. $\alpha$ and $\beta$ are two constants. $c$ is the color parameter, and $X_1$ is the stretch factor. Therefore, the likelihood for SN Ia is defined as
\begin{equation}
\chi_{SN}^2=\Delta^{T}C_{SN}^{-1}\Delta,
\end{equation}
where $\Delta=\mu-\mu_{obs}$ and $C_{SN}$ is the covariance matrix.
\subsection{OHD}
For the OHD in Table \ref{tab:1}, the best-fit values of the model parameters can be determined by a likelihood analysis based on the calculation of
\begin{equation}
 \chi^{2}_H=\sum_{i=1}^{29} \frac{(H(z_{i})-H_{obs}(z_{i}))^2}{\sigma_{H}^{2}(z_{i})}.
\end{equation}
\begin{table}
\centering
\begin{tabular}{|c|c|c|}
\hline
{$z$}   & $H(z)$ &  Ref.\\
\hline
$0.0708$   &  $69.0\pm19.68$         &  Zhang et al. (2014)-\cite{zhang2014four}   \\
    $0.09$       &  $69.0\pm12.0$            &  Jimenez et al. (2003)-\cite{Jimenez2003}   \\
    $0.12$       &  $68.6\pm26.2$           &  Zhang et al. (2014)-\cite{zhang2014four}  \\
    $0.17$       &  $83.0\pm8.0$             &  Simon et al. (2005)-\cite{Simon2005}     \\
    $0.179$     &  $75.0\pm4.0$           &  Moresco et al. (2012)-\cite{Moresco2012}     \\
    $0.199$     &  $75.0\pm5.0$            &  Moresco et al. (2012)-\cite{Moresco2012}     \\
    $0.20$         &  $72.9\pm29.6$         &  Zhang et al. (2014)-\cite{zhang2014four}   \\
    $0.27$       &  $77.0\pm14.0$         &    Simon et al. (2005)-\cite{Simon2005}   \\
    $0.28$       &  $88.8\pm36.6$        &  Zhang et al. (2014)-\cite{zhang2014four}   \\
    $0.352$     &  $83.0\pm14.0$          &  Moresco et al. (2012)-\cite{Moresco2012}   \\
    $0.3802$     &  $83.0\pm13.5$         &  Moresco et al. (2016)-\cite{Moresco2016}   \\
    $0.4$         &  $95\pm17.0$             &  Simon et al. (2005)-\cite{Simon2005}     \\
    $0.4004$     &  $77.0\pm10.2$          &  Moresco et al. (2016)-\cite{Moresco2016}   \\
    $0.4247$     &  $87.1\pm11.2$         &  Moresco et al. (2016)-\cite{Moresco2016}   \\
    $0.4497$     &  $92.8\pm12.9$        &  Moresco et al. (2016)-\cite{Moresco2016}   \\
    $0.4783$     &  $80.9\pm9.0$         &  Moresco et al. (2016)-\cite{Moresco2016}   \\
    $0.48$       &  $97.0\pm62.0$         &  Stern et al. (2010)-\cite{Stern2010}     \\
    $0.593$     &  $104.0\pm13.0$        &  Moresco et al. (2012)-\cite{Moresco2012}   \\
    $0.68$       &  $92.0\pm8.0$        &  Moresco et al. (2012)-\cite{Moresco2012}   \\
    $0.875$     &  $125.0\pm17.0$       &  Moresco et al. (2012)-\cite{Moresco2012}   \\
    $0.88$       &  $90.0\pm40.0$         &  Stern et al. (2010)-\cite{Stern2010}     \\
    $0.9$         &  $117.0\pm23.0$        &  Simon et al. (2005)-\cite{Simon2005}  \\
    $1.037$     &  $154.0\pm20.0$          &  Moresco et al. (2012)-\cite{Moresco2012}   \\
    $1.3$         &  $168.0\pm17.0$        &  Simon et al. (2005)-\cite{Simon2005}     \\
    $1.363$     &  $160.0\pm33.6$          &  Moresco (2015)-\cite{Moresco2015}  \\
    $1.43$       &  $177.0\pm18.0$         &  Simon et al. (2005)-\cite{Simon2005}     \\
    $1.53$       &  $140.0\pm14.0$        &  Simon et al. (2005)-\cite{Simon2005}     \\
    $1.75$       &  $202.0\pm40.0$         &  Simon et al. (2005)-\cite{Simon2005}     \\
    $1.965$     &  $186.5\pm50.4$         &   Moresco (2015)-\cite{Moresco2015}  \\
\hline
\end{tabular}
\caption{\label{tab:1} The current available OHD dataset.}
\end{table}

Finally, we have the total likelihood as
\begin{equation}
   \chi_{tot}^2=\chi_{CMB}^2+\chi_{BAO}^2+\chi_{SN}^2+\chi_{H}^2
\end{equation}

\begin{figure}
  \includegraphics[width=1\textwidth]{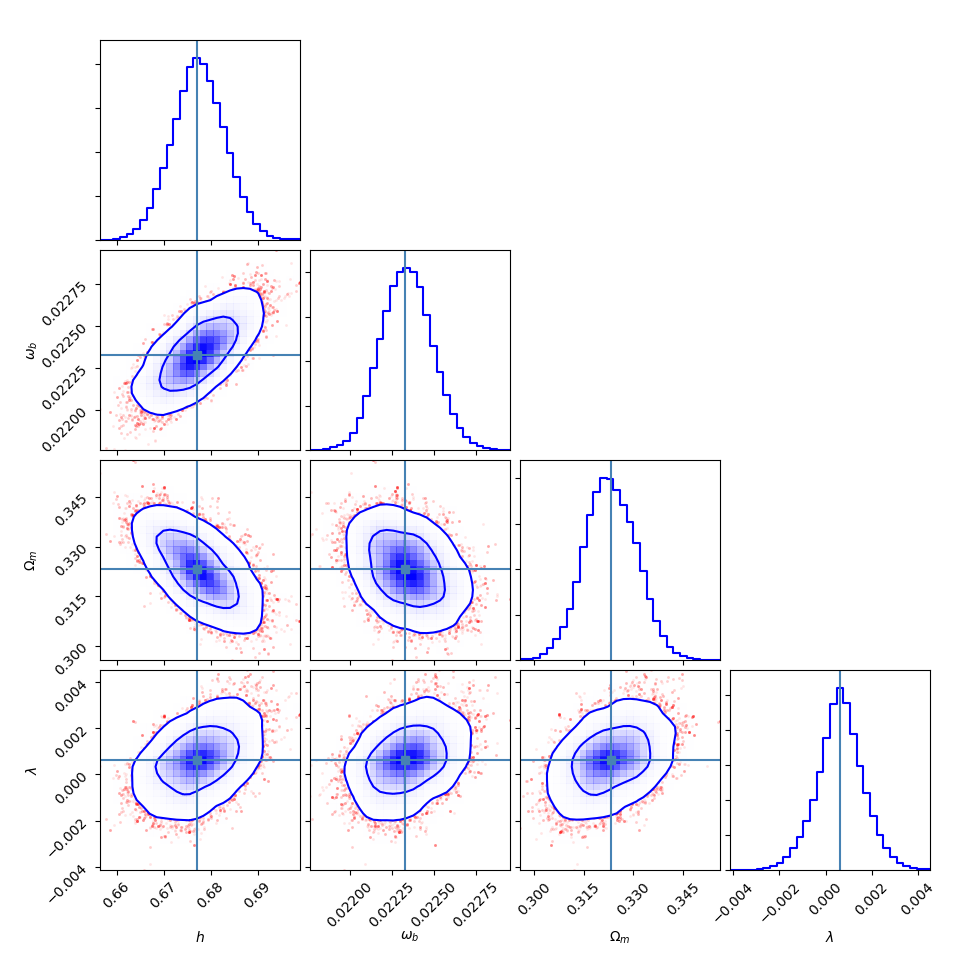}
\caption{$1\sigma$ and $2\sigma$ confidence regions and probability densities for the parameters in the coupled three-form dark energy model.}
\label{fig:1}       
\end{figure}

Since the probability distribution of the parameter $x_{10}$ is approximate to a uniform distribution and its minimum approaches to minus infinity, we set the parameter $x_{10}$ to be fixed at $-1000000$. Therefore, we choose $\Omega_{(m)}$, $\lambda$, $h$, and $\omega_{(b)} $ as fitting parameters for the coupled three-form dark energy model, and $\Omega_{(m)}$, $h$, and $\omega_{(b)}$ as fitting parameters for $\Lambda$CDM. For comparison, we add the $\Lambda$CDM model into our likelihood analysis. The fitting results are presented in Tab.\ref{tab:2} and Fig.\ref{fig:1}. From Tab.\ref{tab:2}, we obtain $h=0.678_{-0.006}^{+0.007}$($1\sigma$ level) km/s/Mpc for the coupled three-form dark energy model, and $h=0.676_{-0.005}^{+0.005}$($1\sigma$ level) km/s/Mpc for the $\Lambda$CDM model. Corresponding, the $H_0$ tension between them and $h=0.738\pm0.011$ are reduced to $4.3\sigma$ for the coupled three-form dark energy model, and $5.1\sigma$ for the $\Lambda$CDM model. Therefore, within the coupled three-form dark energy model, the tension is still strong. As shown in Fig.\ref{fig:1}, considering the extra parameters $\lambda$ in the $\Lambda$CDM model can affect the constraints on the Hubble constant $H_0$ because that $\lambda$ are positively correlated with $h$. These fitting results are not unexpected, in fact, serval recent works on coupled dark energy model \cite{di2020nonminimal,di2020interacting,cheng2020testing} produce similar results, i.e. $H_0$ gets a bit higher but not enough to completely solve the $H_0$ tension, which is mostly alleviated by a bit larger error bars.

We also consider the Akaike information criterion (AIC) to compare the coupled three-form dark energy model and the $\Lambda$CDM model. The AIC is defined as $\chi_{min}^2 + 2k$, where $k$ denotes the number of cosmological parameters. Therefore, from Tab.\ref{tab:2}, we have AIC=$715.734$ for the coupled three-form dark energy model and AIC=$713.873$ for the $\Lambda$CDM model. Then the $\Lambda$CDM model is more supported than the coupled three-form dark energy model because that, by definition, a model with a smaller value of AIC is a more supported model.
 \begin{table}
\begin{center}
\begin{tabular}{|c|c|c| }
\hline Model &coupled three-form dark energy model &   $\Lambda$CDM
\\ \hline
$\Omega_{(m)}$    &$0.323 _{-0.007}^{+0.008}$ &  $ 0.321_{-0.007}^{+0.007}$
                     \\
$\lambda$         &  $0.001_{-0.001}^{+0.001}$ & $-$
                     \\
 $h$          &    $ 0.679_{-0.007}^{+0.008}$ & $0.676_{-0.005}^{+0.005}$
                       \\
 $\omega_{(b)}$          &    $ 0.0223_{-0.0002}^{+0.0002}$ &  $0.0223_{-0.0001}^{+0.0001}$
                       \\
\hline
$\chi_{min}^2$ & 707.734  & 707.873
  \\
AIC & 715.734   & 713.873
\\
\hline
\end{tabular}
\caption{ Fit results of cosmological parameters in the coupled three-form dark energy model and $\Lambda$CDM using the CMB+BAO+JLA+OHD data.}
\label{tab:2}
\end{center}
\end{table}

\section{Conclusions}
In this paper, a coupled three-form dark energy model is put forward to relieve the Hubble tension. To start with, we perform a dynamical analysis with the coupled three-form dark energy model, and obtain three fixed points, including a saddle point representing a radiation dominated Universe, a saddle point representing a matter dominated Universe, and a attractor representing a saturated de Sitter Universe. Then, we confront the $\Lambda$CDM model and the coupled three-form dark energy model with CMB, BAO, SN Ia observation, and obtain $h=0.676_{-0.005}^{+0.005}$($1\sigma$ level) km/s/Mpc for the $\Lambda$CDM model, which is in strong tension with latest local $h$ value at $5.1\sigma$ level, and $h=0.679_{-0.007}^{+0.008}$($1\sigma$ level) km/s/Mpc for the coupled three-form dark energy model, which is in strong tension with latest local $h$ value at $4.3\sigma$ level. Therefore, within the coupled three-form dark energy model, the strong $H_0$ tension is still exist. In addition, by using AIC, we find that the $\Lambda$CDM model is more supported than the coupled three-form dark energy model.
\section*{Acknowledgments}
The paper is partially supported by the Natural Science Foundation of China.

\end{document}